\documentclass[runningheads]{llncs}

\usepackage[square,sort,comma,numbers,sectionbib]{natbib}
\usepackage{graphicx}
\usepackage{subcaption}
\usepackage{multirow}
\usepackage{hyperref}
\usepackage{booktabs}

\PassOptionsToPackage{table}{xcolor}
\usepackage{xcolor}

\providecommand{\myparab}[1]{\smallskip\noindent\textbf{#1} }

\renewcommand{\footnotesize}{\scriptsize}
\newcommand{\sectionref}[1]{\textsection\ref{#1}}
\providecommand{\sysname}{\textit{\texttt{DNEye}}}
\providecommand{\myparab}[1]{\smallskip\noindent\textbf{#1} }

\begin{document}
\title{Measuring the Accessibility of Domain Name\\Encryption and Its Impact on Internet Filtering}
\titlerunning{Domain Name Encryption and Its Impact on Internet Filtering}
\author{Nguyen Phong Hoang,\inst{1}
Michalis Polychronakis,\inst{2}
Phillipa Gill\inst{3}}
\authorrunning{NP Hoang et al.}
%
\institute{University of Chicago,
\email{nguyenphong@uchicago.edu}\\ \and
Stony Brook University,
\email{mikepo@cs.stonybrook.edu}\\ \and
Google Inc.,
\email{phillipagill@google.com}}

\maketitle              

\begin{abstract}
Most online communications rely on DNS to map domain names to their hosting IP
address(es). Previous work has shown that DNS-based network interference is
widespread due to the unencrypted and unauthenticated nature of the original DNS
protocol. In addition to DNS, accessed domain names can also be monitored by
on-path observers during the TLS handshake when the SNI extension is used. These
lingering issues with exposed plaintext domain names have led to the development
of a new generation of protocols that keep accessed domain names hidden.
DNS-over-TLS (DoT) and DNS-over-HTTPS (DoH) hide the domain names of DNS
queries, while Encrypted Server Name Indication (ESNI) encrypts the domain name
in the SNI extension.

We present \sysname, a measurement system built on top of a network of
distributed vantage points, which we used to study the accessibility of DoT/DoH
and ESNI, and to investigate whether these protocols are tampered with by
network providers (e.g., for censorship). Moreover, we evaluate the efficacy of
these protocols in circumventing network interference when accessing content
blocked by traditional DNS manipulation. We find evidence of blocking efforts
against domain name encryption technologies in several countries, including
China, Russia, and Saudi Arabia. At the same time, we discover that domain name
encryption can help with unblocking more than 55\% and 95\% of censored domains
in China and other countries where DNS-based filtering is heavily employed.

\end{abstract}

\section{Introduction}

Despite its importance, the domain name system (DNS)~\cite{rfc1034} was not
designed with encryption or authentication. Traditional DNS resolutions are
transmitted in plaintext, allowing network-level adversaries to easily eavesdrop
or tamper with the resolution process~\cite{pearce2017global, niaki2020triplet,
GFWatch}, jeopardizing user privacy and security.

Additionally, the domain name information is also visible in the Transport Layer
Security (TLS) protocol~\cite{rfc3546}. During the TLS handshake, the client
specifies the domain name in the Server Name Indication (SNI) in
plaintext~\cite{rfc5246}, signaling a server that hosts multiple domain names
(name-based virtual hosting) to present the correct TLS certificate to the
client. However, network observers can also use this information to surveil or
interfere with a user's connection.

With the proliferation of network interference and Internet
surveillance~\cite{Fuchs2011InternetAS}, users have become more aware of their
online security and privacy. This has led to DNS and TLS improvements for
enhancing user privacy. DNS-over-TLS (DoT)~\cite{rfc7858}, DNS-over-HTTPS
(DoH)~\cite{rfc8484}, and Encrypted Server Name Indication
(ESNI)~\cite{tlsdraft05} are recently proposed privacy-enhancing protocols, to
which we refer collectively as domain name encryption technologies.

However, advances in domain name encryption technologies have not gone unnoticed
to censors. For instance, China has been blocking ESNI since July
2020~\cite{esnichina}. Russia has also drafted laws to ban the adoption of
domain name encryption~\cite{russiaesni}. Despite these reports, there has yet
to be a comprehensive study to shed light on how common blocking of domain name
encryption is; and whether domain name encryption approaches can help with
evading network interference.

In this paper, we present \sysname, a measurement system built on top of a
network of vantage points, allowing us to study the accessibility of domain name
encryption technologies and whether censors are interfering with them, and to
evaluate their efficacy in bypassing network interference. Over a period of six
months, \sysname\ conducted 315K measurements to examine the accessibility of
1.6K domains and DoT/DoH (hereafter: DoTH) resolvers around the globe.

While our data shows that DNS manipulation is widespread, we found no major
DNS-based filtering of DoTH resolvers' domain names at the autonomous system
(AS) level, except for \textit{ordns.he.net}, which is blocked by the Great
Firewall via DNS poisoning, and two Cloudflare servers
(\textit{cloudflare-dns.com} and \textit{mozilla.cloudflare-dns.com}) blocked in
Thailand's AS23969. We then examine whether connections destined for DoTH
resolvers suffer from any interference (\sectionref{sec:dne_connectivity}). We
detect several ASes in China interfere with connections destined for different
DoTH resolvers. We also found only 1.5--2.25\% of the domains in the top-level
domain (TLD) zone files with ESNI supported. Despite this small number of
ESNI-supported domains, we find evidence that China and numerous network
operators in Russia have started blocking connections to ESNI-enabled websites
(\sectionref{sec:dne_connectivity}).

Finally, we investigated whether domain name encryption can help with bypassing
Internet filtering (\sectionref{sec:circumvention}), and found that it can help
with unblocking many censored domains. Specifically, except from Iran, we could
successfully fetch more than 55\% and 95\% of the blocked domains in China and
other countries where DNS-based network filtering is widely employed.
\section{Background}
\label{sec:background}

\subsection{Common Internet Filtering Techniques}
\label{sec:InternetFiltering}


There are several Internet filtering techniques often employed by authoritarian
governments to control the free flow of information.

\myparab{DNS manipulation.} Due to the unencrypted design of the original DNS
protocol~\cite{rfc1034}, any on-path network observer can monitor the domain
name being queried by a user. The visibility into the plaintext domain name
allows any on-path filtering system to trivially conduct DNS-based filtering.
Specifically, an on-path observer can forge DNS responses containing
non-routable IPs, an IP under its control, or a DNS error code. China's Great
Firewall (GFW) is one of the most prominent filtering systems that injects such
forged DNS packets in response to ``sensitive'' DNS
queries~\cite{Anonymous2014a,niaki2020triplet,GFWatch}.

\myparab{IP blocking.} Once a user obtains the correct IP(s) of the intended
website, a TCP connection is established with the web server for data
transmission. Upon observing a connection attempt to a forbidden IP, filtering
systems often inject RST (reset) packets to interfere with the TCP
stream~\cite{Wang2017YourSI, Bock2019GenevaEC, hoang:2019:measuringI2P}. In
other cases, null routing~\cite{rfc3882} can also be used to discard traffic
destined for certain IPs.

\myparab{Application-level interference.} After establishing the TCP connection,
the user proceeds with sending an HTTP request with the HTTP Host field
specifying the intended domain name. Similarly, for HTTPS-supported websites,
clients specify the intended domain name in the SNI field of the TLS
handshake~\cite{rfc5246}. Filtering systems can also monitor these fields to
determine the domain name being visited to interfere with the connection, either
by injecting RST packets or modifying the HTTP traffic to redirect the user to a
blockpage~\cite{Jones2014AutomatedDA, hoang:2019:measuringI2P, niaki2020iclab,
sundara2020censored}.

\subsection{Domain Name Encryption Protocols}
\label{sec:dne}

As discussed in~\sectionref{sec:InternetFiltering}, the exposure of the domain
information in both DNS and TLS protocols has been widely exploited for network
interference~\cite{GFWatch, pearce2017global,hoang:2019:measuringI2P,chaiesni}.

\myparab{Encrypted DNS.} DoT~\cite{rfc7858} and DoH~\cite{rfc8484} were proposed
to provide integrity and confidentiality for DNS resolutions by encrypting DNS
packets between clients and DoTH resolvers. These protocols have been
standardized and supported by many major Internet companies.
Google~\cite{googleDoH} and Cloudflare~\cite{cloudflare_DoT} have provided
public DoTH resolvers, while popular web browsers (e.g,
Firefox~\cite{firefoxDoH_start}, Chrome~\cite{chrome_doh},
Safari~\cite{appleDoH_start}, and Edge~\cite{edgeDoH_start}) have also supported
DoH.

\myparab{Encrypted SNI.} The SNI extension~\cite{rfc3546} was introduced to
enable name-based virtual hosting. Up until TLS 1.2~\cite{rfc5246}, clients
indicate their intended domain name in the SNI field during the TLS handshake in
plaintext so that the server can present the appropriate certificate. Encrypted
SNI is one of the optional extensions of TLS 1.3 designed to conceal the domain
name information~\cite{tlsdraft02}. ESNI has been reworked to Encrypted Client
Hello (ECH)~\cite{tlsdraft07} since June 2020.

\section{\sysname\ Design}
\label{sec:method}

\begin{figure*}[t]
  \centering
  \includegraphics[width=.8\columnwidth]{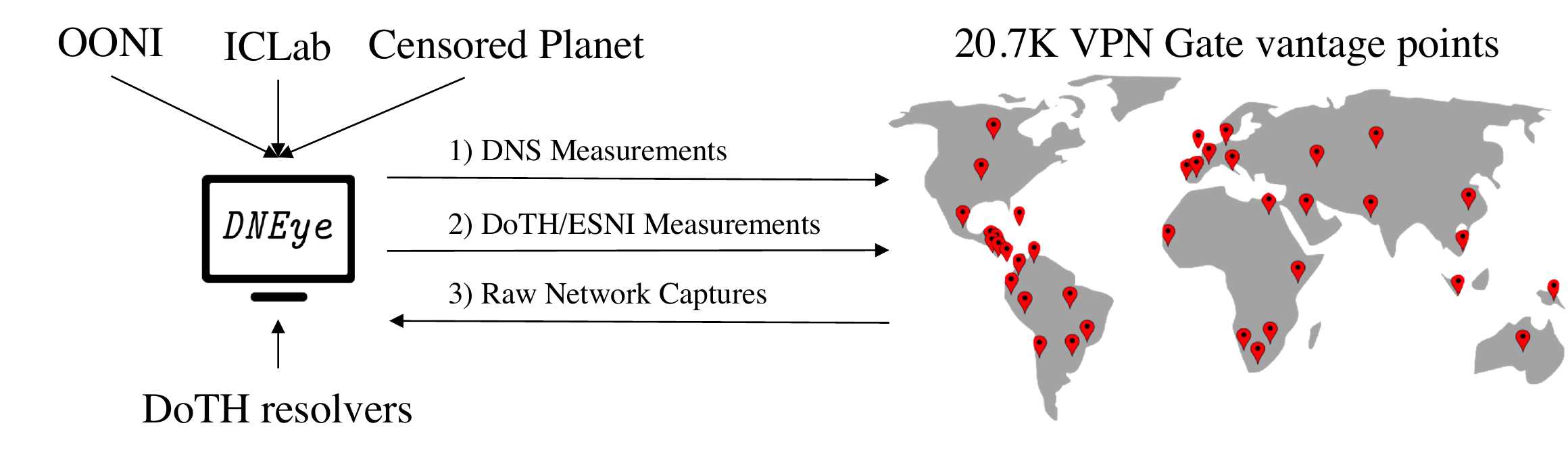}
  \caption{\sysname\ architecture.}
  \label{fig:dneye}
\end{figure*}

Given that the visibility into plaintext domain information is lost due to the
introduction of domain name encryption, we are interested in investigating how
these new protocols impact Internet filtering systems. We developed \sysname\ to
(1) assess the current situation of DNS-based network filtering, (2) examine the
accessibility of domain name encryption protocols and whether they are
interfered with across different network locations, and (3) evaluate whether
these protocols can help with evading network interference. In this section, we
first describe how we obtain testing vantage points, their limitations, and
ethical considerations. We then explain the process by which we curate our test
list of domains and how we use \sysname\ to perform various connectivity
measurements. Figure~\ref{fig:dneye} provides an overall view of \sysname's
architecture.

\subsection{Vantage Points}
\label{sec:vantages}

\myparab{Vantage points.} The core component of \sysname\ is a network of
vantage points (VPs) provided by VPN Gate~\cite{nobori2014vpn}. VPN Gate is a
public VPN relay service where any volunteer can register to be a VPN endpoint
by running the SoftEther package~\cite{SoftEther}. Since these VPs are operated
by volunteers around the globe, they often reside in residential networks,
allowing us to observe filtering policies which would be usually unobservable
via commercial VPNs in data centers~\cite{hoang:2019:measuringI2P}. However,
since the infrastructure is volunteer-based, these VPs are often short-lived and
unsuitable for testing a long list of websites. We describe how we account for
this shortcoming by conducting a sliding-window analysis
in~\sectionref{sec:dns_filtering}.

Table~\ref{table:vantagepoints} summarizes the geographical distribution of our
VPs by continent and Internet freedom scores assessed by the Freedom
House~\cite{freedom}. During six months of our study, VPN Gate VPs provide us
with access to 34K unique IPs. We however exclude 13.3K short-lived VPs that
were online for less than one day to eliminate unstable data points. In total,
\sysname\ has access to about 20.7K VPs in 85 countries, with an average of 10
ASes per country. More importantly, we have access to more than twice the number
of countries classified as ``not free'' than ICLab~\cite{niaki2020iclab}, which is
a platform that also relies on VPN services for measuring network interference.
Of these ``not free'' countries, there are 11 countries where we have access to
VPs located in at least two different ASes, allowing us to observe centralized
country-level filtering policies (if any).

\begin{table}[t]
  \caption{Geographical distribution of \sysname's VPN vantage points indicated
    by the number of countries and ASes across continents. \texttt{NF},
    \texttt{PF}, and \texttt{F} columns denote the number of \emph{politically
    not free}, \emph{partially free}, and \emph{free} countries.}
  \label{table:vantagepoints}
  \centering
    \small
    \begin{tabular}{lrrrrrr}
    \toprule
    \scriptsize\bfseries Continent &
    \scriptsize\bfseries Vantages &
    \scriptsize\bfseries Countries &
    \scriptsize\bfseries ASes &
    \scriptsize\bfseries NF &
    \scriptsize\bfseries PF &
    \scriptsize\bfseries F \\
    \midrule
    Asia       &  14K    &  32/48          &  367  & 17 &  8  &  6  \\
    Africa     &  13     &  4/54           &  9    & 2  &  0  &  2 \\
    N. America &  2.7K   &  6/23           &  157  & 0  &  1  &  3 \\
    S. America &  811    &  9/14           &  58   & 1  &  4  &  4 \\
    Europe     &  2.8K   &  32/50          &  271  & 2  &  3  & 27 \\
    Oceania    &  282    &  2/\phantom{0}6 &  16   & 0  &  0  &  2 \\
    \addlinespace
    Total      & 20.7K   &  85/195         & 878  & 22  & 16 &  44\\
    \bottomrule
    \end{tabular}
    \vspace{-0.3cm}
\end{table}

\myparab{Ethical considerations.} Measuring Internet interference using
volunteer-based VPs must be carried out in a thoughtful way that takes into
consideration various ethical aspects~\cite{jones2015ethical}. There are
commercial services (e.g., Luminati~\cite{luminati}) that provide residential
VPs, meeting the measurement needs of our study. Nonetheless, studies have
reported illicit activities of these services, e.g., malware
hosting~\cite{mi2019resident}. We, thus, opt to use VPN Gate for two primary
reasons. First, to become a VPN server, the SoftEther VPN
package~\cite{SoftEther} requires all volunteers to manually go through a
process that reminds them about the associated risks of joining the VPN Gate
research network~\cite{vpngateprocess}. Therefore, it is reasonable to expect
volunteers, who are willing to be VPN endpoints, to fully understand the
potential risks before agreeing to share their network connection. Moreover, the
University of Tsukuba and the VPN Gate software also record access logs which
serve as an anti-abuse policy used by the project and to assist its volunteers
in case of disputes~\cite{vpngate_disclosure}. Since the launch of \sysname\ we
have not received any complaints.

\subsection{Test List}
\label{sec:testlist}
While it is desirable to test many domains, due to the short-lived nature of
VPs, we cannot test a large number of domains. OONI~\cite{filasto2012ooni},
ICLab~\cite{niaki2020iclab}, and Censored Planet~\cite{sundara2020censored} are
measurement platforms actively monitoring Internet interference around the
globe. Since these platforms have implemented testing modules to monitor DNS
filtering, we opt to use their collected data as an input for \sysname.

We first look for domains reported as censored by these platforms within the
past 30 days and visit them to confirm their online status. We consider domains
that are censored in at least two ASes per country and reported by at least two
platforms. This helps eliminate unreliable data points that could have been
caused by generic network errors. To that end, we obtain 1.5K domains that are
commonly reported as censored by these prior platforms in 77 countries where we
have VPs. Since one of our main goals is to examine the accessibility of domain
name encryption protocols, we also add domains for 71 DoTH resolvers publicly
available at the time of our study~\cite{dohservers}. These resolvers are
indexed in Table~\ref{tab:doth_mapping}.

\subsection{Measurements}
\label{sec:measurements}

Once connected to a VP, we instruct \sysname\ to capture all network traffic
during each measurement. Monitoring traffic transmitted over the VPN tunnel
enables us to observe network interference (if any) across all layers of the
network stack.

\myparab{Measuring DNS manipulation.} After each VPN tunnel is established,
\sysname\ first issues DNS queries for the domains in our test list. This allows
us to not only obtain an updated view of DNS filtering across network locations,
but also determine whether there are any filtering systems that block these DoTH
resolvers via DNS tampering. \sysname\ sends DNS queries to both public DNS
resolvers (e.g., Google and Cloudflare) and the local DNS resolver configured by
each VP's network provider. Querying both types of resolvers helps us discern
whether DNS tampering (if any) is conducted solely by the local resolver or by
an on-path system between our clients and the selected public resolvers.

\myparab{Measuring DoTH and ESNI connectivity.} \sysname\ then uses
\texttt{kdig}~\cite{kdig} to send encrypted DNS queries to 71 DoTH resolvers to
resolve a control domain for which we know the correct answer. This test checks
whether each DoTH server returns the control domain's correct IP. The ability to
capture network packets allows us to detect at which stage of a connection
(i.e., TCP or TLS handshakes) a filtering system tampered with the connection
destined for the selected DoTH servers. In addition, to determine whether ESNI
is blocked, \sysname\ also attempts to connect to an ESNI-supported website
under our control.

\myparab{Measuring filtering circumvention.} Finally, to evaluate whether domain
name encryption can help evade Internet filtering, \sysname\ instruments a
customized web browser with DoH and ESNI enabled to crawl filtered domains from
VPs where DNS filtering of these domains was observed in the first step.

However, as later shown in~\sectionref{sec:dne_connectivity}, many DoTH
resolvers are being blocked in several countries. To prevent any filtering
system from interfering with our DoH resolutions, we configure the crawler to
use our private DoH resolver, which runs on a non-standard port (i.e., different
from 443) and is hosted in an uncensored network. For an ESNI-supported website,
the crawler will also obtain its ESNI key and establish an ESNI-enabled TLS
connection. Simultaneously, we crawl the same website from an uncensored control
environment for later comparison.

Between November 12, 2020, and May 12, 2021, \sysname\ has conducted 315K
connectivity measurements for 1.6K domains and DoTH resolvers in 878 ASes across
85 countries. The aggregated dataset will be made available to the public to
stimulate future studies in this domain at
\url{https://homepage.np-tokumei.net/publication/publication_2022_pam/}.
\section{Results}
\label{sec:analysis}

\subsection{DNS-based Network Interference}
\label{sec:dns_filtering}

To identify DNS tampering, we apply well-established heuristics in the
literature on the data collected by \sysname\
(Appendix~\ref{appendix:dns_tampering}). For each DNS query sent via a VP, we
extract all DNS responses captured from that VP's network traffic. In case of a
poisoned DNS response, the ability to analyze raw network packets allows us to
discern whether it was injected by an on-path filtering system or directly
served from the local DNS resolver. Specifically, if the tampering is conducted
by an on-path filtering system when querying a public DNS resolver, we will be
able to observe more than one DNS response, of which the one arrives the VP
first is usually forged by the filtering system~\cite{GFWatch, duan2012hold}. In
case of a forged response served directly from a local resolver, we will only
observe that one response.

\myparab{Sliding-window analysis.} Due to the short-lived nature of our VPs, we
do not have access to all VPs on the same day. To reduce the impact of
unreliable data points, we analyze the data by considering a sliding window of
seven days for each measurement. In other words, for each domain tested in a
measurement, we aggregate the data we have from the same VP within a window of
±3 days. Meanwhile, we also compute the average filtering rate (i.e., the number
of measurements we mark as ``tampering'' divided by the total measurements for
each VP and domain pair). If the filtering rate of a domain at a VP is higher
than 80\%, we consider that domain as ``blocked'' at that VP in that particular
measurement. We conservatively choose the 80\% threshold to avoid false
positives caused by generic network errors instead of network interference.

\begin{table}
  \vspace{-0.8cm}
  \centering
  \caption{Top five countries where most DNS resolutions are tampered with.}
  \label{tab:dns_table}

  \begin{subtable}[c]{0.48\columnwidth}
    \subcaption{When querying local resolvers}
    \label{tab:dns_table_local}
    \centering
    \begin{tabular}{l@{\hskip 0.2in}r}
      \toprule
      \textbf{Country} &\textbf{Domains}   \\ [0.5ex]
      \midrule
      China     &     305 \\
      Russia    &     251 \\
      Japan     &     181  \\
      Iran      &     159  \\
      Indonesia &     135  \\
      \bottomrule
    \end{tabular}
  \end{subtable}
  ~
  \begin{subtable}[lr]{0.48\columnwidth}
    \subcaption{When querying public resolvers}
    \label{tab:dns_table_public}
    \centering
    \begin{tabular}{l@{\hskip 0.2in}r}
      \toprule
      \textbf{Country} &\textbf{Domains}   \\ [0.5ex]
      \midrule
      China     &     300 \\
      Russia    &     205 \\
      Iran      &     147  \\
      Indonesia &     134  \\
      India     &     98  \\
      \bottomrule
      \end{tabular}
  \end{subtable}
  \vspace{-0.4cm}
\end{table}

Regardless of the introduction of DNS encryption protocols, our results confirm
that DNS manipulation is still widely employed, aligning with prior
reports~\cite{pearce2017global, niaki2020iclab, sundara2020censored,
jin2021understanding, GFWatch}. Table~\ref{tab:dns_table} presents the top five
countries where most DNS resolutions are tampered with. Japan is not a
censorship country, as is evident by its ``\emph{free}'' classification by the
Freedom House~\cite{freedom}. The reason for the high number of DNS resolutions
interfered by local DNS resolvers is that VPN Gate is a Japan-based project,
thus providing us with a large number of VPs from many residential networks
across Japan. Our collected data indicates that many of these VPs are configured
with filtering services provided by local DNS resolvers. Hence, queries sent to
these resolvers are often interfered with. More specifically, many DNS responses
returned by these local DNS resolvers contain IPs redirecting to destinations
within the same AS of the VPs where DNS queries were issued. On the other hand,
we could still obtain the correct DNS records for our DNS resolutions when
querying public resolvers from these same VPs. As a result, this is not a case
of country-level DNS censorship.

\myparab{DNS manipulation of DoTH domains.} As laid out
in~\sectionref{sec:measurements}, \sysname\ also performs DNS resolutions for
domain names of 71 DoTH resolvers to determine whether there is any DNS
tampering against these resolvers. Except for China, we did not observe any
DNS-based filtering of DoTH domain names at country level. Specifically,
\texttt{ordns.he.net} is blocked in China by the GFW via DNS tampering. In
addition, \sysname\ detected DNS tampering against two Cloudflare servers
(\textit{cloudflare-dns.com}, \textit{mozilla.cloudflare-dns.com}) by AS23969
TOT Public Company Ltd, Thailand. DNS resolutions for these two domains are
poisoned with a forged IP (i.e., 180.180.255.130), pointing to a blockpage.

\subsection{DoTH and ESNI Accessibility}
\label{sec:dne_connectivity}

\myparab{DoTH accessibility.} Since DoTH is still in its early stage of adoption
while not all DoTH servers are well-provisioned, any of them may become
unavailable during our measurement, e.g., due to maintenance. To determine if a
resolver is unavailable due to a generic reason rather than network
interference, we aggregate all daily resolutions from all VPs for that
particular resolver. If more than 70\% of the queries were successfully
resolved, we consider that resolver as available on that day. We choose 70\% as
a conservative threshold to prevent intermittently available resolvers from
causing false positives in our analysis.

\begin{figure*}[t!]
  \centering
    \begin{subfigure}[t]{0.48\columnwidth}
      \centering
      \includegraphics[width=\columnwidth]{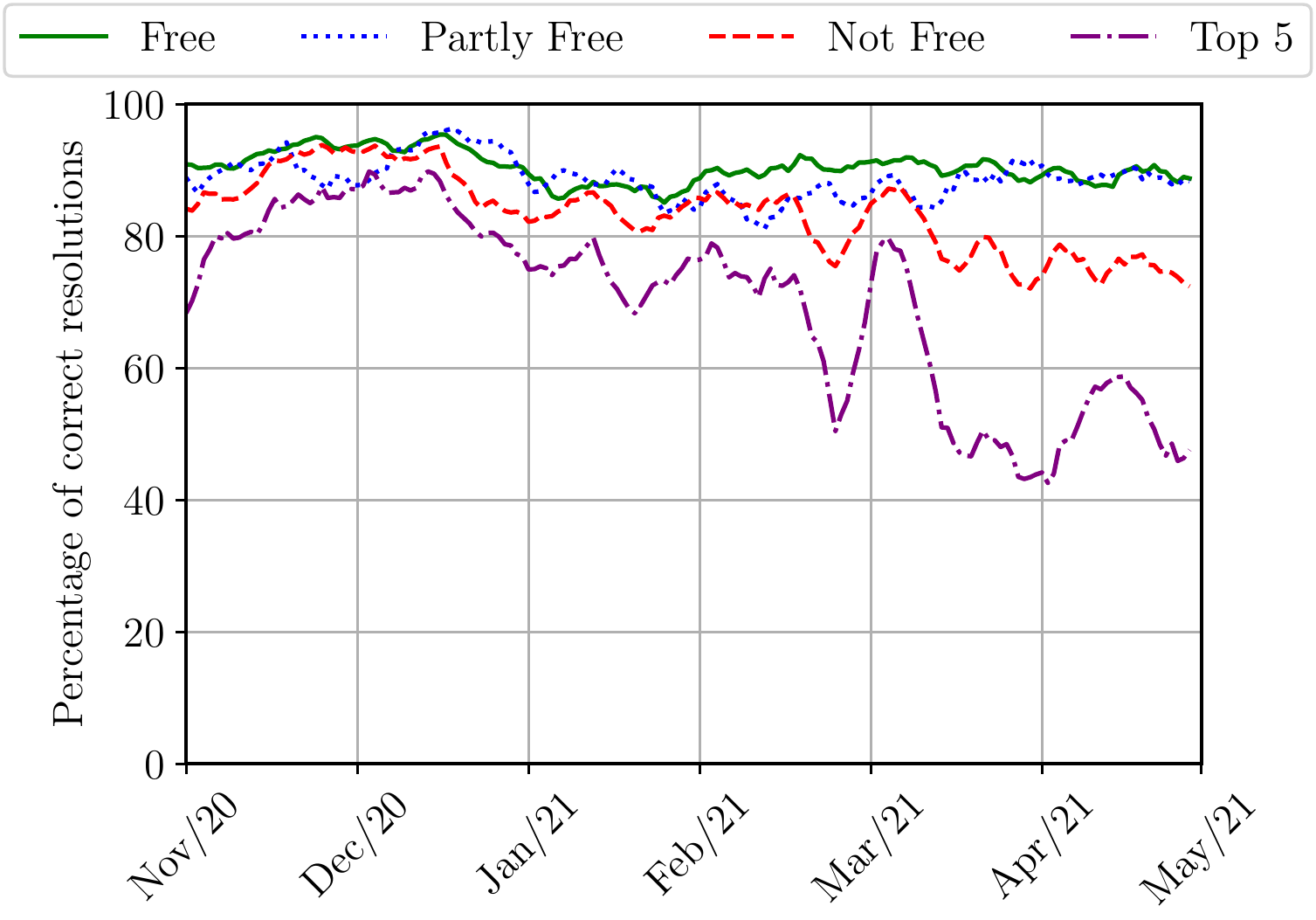}
      \caption{Resolutions using DoT}
      \label{fig:dot_accessibility}
    \end{subfigure}
  ~
    \begin{subfigure}[t]{0.48\columnwidth}
      \centering
      \includegraphics[width=\columnwidth]{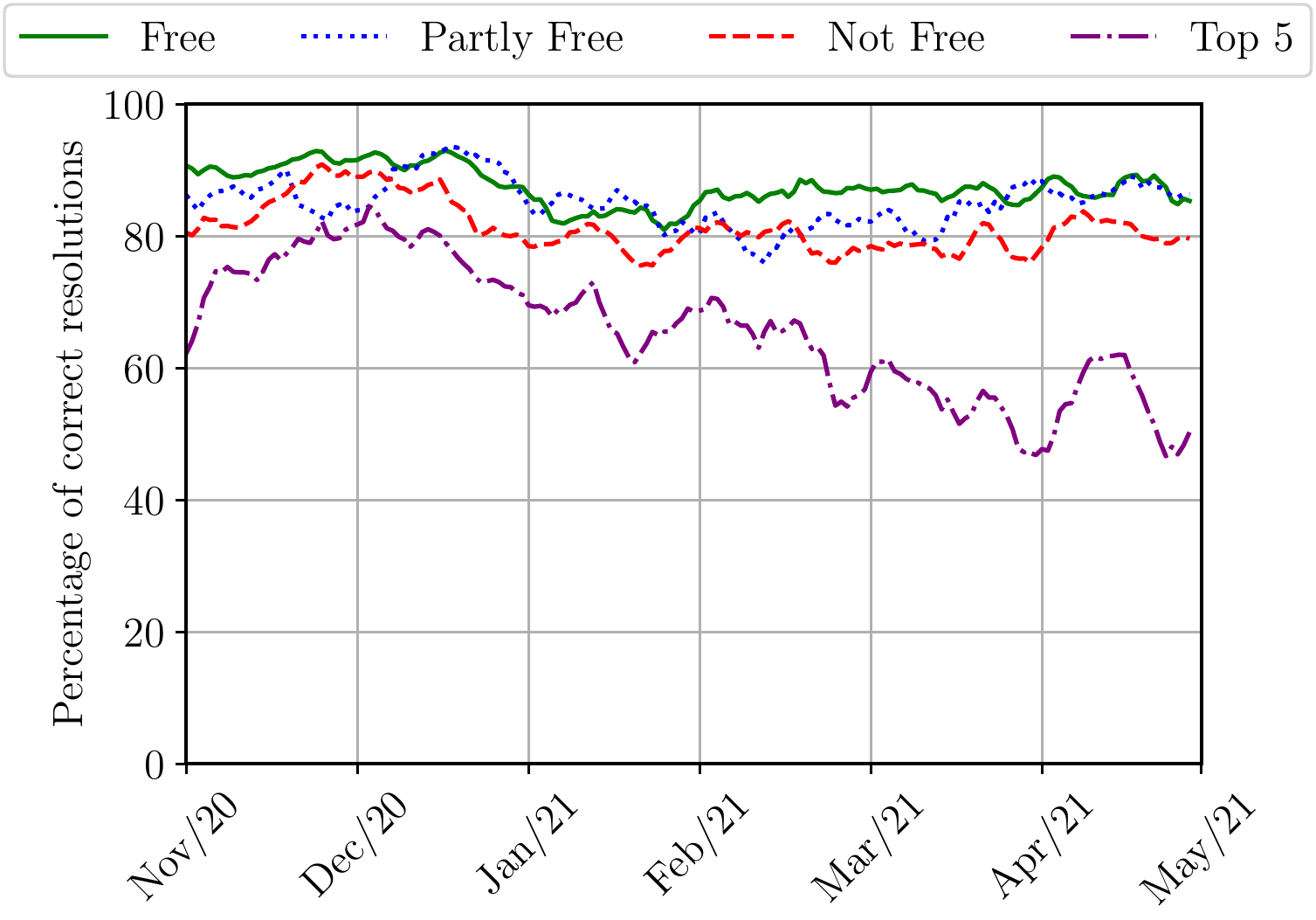}
      \caption{Resolutions using DoH}
      \label{fig:doh_accessibility}
    \end{subfigure}%
  \caption{Percentage of correct DoTH resolutions over time.}
  \label{fig:doth_accessibility}
\end{figure*}

Figures~\ref{fig:dot_accessibility} and~\ref{fig:doh_accessibility} show the
percentages of correct resolutions performed daily using DoT and DoH resolvers,
respectively. We consider a resolution as correct when the correct IP of our
control domain is successfully returned. The result is clustered by country type
defined by the Freedom House~\cite{freedom}. The percentages of correct
resolutions obtained via VPs in \emph{``not free''} countries are lower than
those in \emph{``partially free''} and \emph{``free''} countries. To better
highlight this finding, we add to both plots another dash-dot (purple) line,
computed from data of the top five \emph{``not free''} countries that have the
most number of failed resolutions, namely China, Russia, Iran, Saudi Arabia, and
Venezuela. It is visible on the plots that the number of successful resolutions
for these five countries has decreased significantly since March. This decrease
is driven by the blocking effort of China, where our system detected an increase
in network interference with our DoTH resolutions issued from China VPs. In an
earlier study, Lu et al.~\cite{Lu2019AnEL} reported successful rates of more
than 84\% and 99\% for resolutions using Cloudflare and Quad9 DoT resolvers,
respectively. However, since early March, \sysname\ has detected increasing
network interference efforts by the GFW against DoT resolutions destined for
several major resolvers, including Cloudflare, Quad9, AdGuard, and
CleanBrowsing. Our findings corroborate several anecdotal reports from users in
China about DoTH blocking around that same time~\cite{chinablocksDoTH}.

\myparab{DoTH filtering.}
To examine how filtering systems interfere with connections destined for DoTH
resolvers, we analyze packets captured by \sysname\ for measurements in which
failed resolutions were observed. The ability to process raw network packets
allows us to pinpoint the stage at which a connection was interfered with, thus
being able to identify the employed filtering technique (i.e., TCP packet
injection, SNI-based filtering, or packet dropping).

We employ the same sliding-window technique defined
in~\sectionref{sec:dns_filtering} to determine blocking cases of DoTH resolvers.
Given a VP, an average failure rate of a DoTH resolver is calculated by dividing
the number of failed resolutions by the total number of resolutions performed at
that VP within a seven-day window. If the failure rate exceeds 80\%, we label
the DoTH resolver as ``probably blocked'' at that VP. To determine whether a
DoTH resolver is actually blocked at a VP, we then compute the 90th percentile
value of all failure rates for that DoTH resolver at that particular VP. If the
value is greater than 80\%, we can be confident that there is network
interference with connections destined for that resolver at that VP. We employ
this 90th percentile threshold in combination with the sliding-window analysis
to account for failed resolutions caused by sporadically available VPs and
unstable DoTH resolvers rather than actual network interference.

Next, we consider a DoTH resolver to be blocked by an AS if \sysname\ detects
network interference from at least two VPs from different subnets of that AS on
two separate days. Table~\ref{tab:AS_doth_filtering} in
Appendix~\ref{appendix:AS_doth_filtering} depicts the top countries where most
AS-level DoTH filtering was detected. China has the most number of ASes that
interfere with DoTH connections. The filtering of different DoTH resolvers
detected at different ASes indicates that DoTH filtering is implemented by
individual Internet service providers rather than a centralized policy (e.g.,
centralized DNS-based filtering by the GFW~\cite{GFWatch}).

Another advantage provided by the VPN Gate VPs is that having access to multiple
ASes per country allows \sysname\ to identify cases of country-level filtering
where multiple ASes interfere with the same domains. For instance, \sysname\
detects SNI-based network interference against the same set of DoH resolvers
across different ASes in Saudi Arabia, indicating a centralized filtering
policy.

In Iran, we also observe filtering of multiple DoTH resolvers. Notably,
SNI-based filtering of TLS connections, destined for both DoT and DoH servers of
\texttt{dns.google}, were detected from several subnets of AS58224. This same
filtering was also detected at AS39501 and AS56402, which are not considered in
Table~\ref{tab:AS_doth_filtering} since we did not have VPs from more than one
subnet in these two ASes.

Filtering of some DoTH resolvers was also detected in the US, South Korea, and
Singapore (Table~\ref{tab:AS_doth_filtering}). However, upon verifying the
organization information of the filtering ASes, we find that these are filtering
cases implemented by corporate and institutional firewalls instead of a
country-wide filtering policy.

\myparab{ESNI adoption.}
For ESNI to provide any meaningful privacy and filtering resistance benefits, it
needs to be supported by many websites, since if there are only a few
ESNI-supported websites, connections to their servers are trivially
distinguishable~\cite{Hoang2020:ASIACCS, Hoang2021:PoPETS}. Therefore, we first
measure the adoption of ESNI on the web by looking up the ESNI TXT record for
more than 350M domains from TLD zone files~\cite{ICANN}.

Over the course of our measurement period, we find that only about 1.5--2.25\%
of domains from TLD zone files have a valid ESNI key format (see
Appendix~\ref{appendix:esni_prevalece}). Of these ESNI-supported domains, 15.4K
and 143.3K domains are within the top 100K and 1M popular domains ranked by the
Tranco list~\cite{LePochat2019}, respectively. We have also measured the
deployment of Encrypted Client Hello (ECH) by probing for HTTPS resource
records~\cite{tlsdraft07} but did not find any evidence of ECH deployment in the
wild.

\myparab{ESNI filtering.} As described in~\sectionref{sec:measurements},
\sysname\ also measures the filtering of ESNI by visiting our control website
that has ESNI enabled. This website will reflect the visiting client's IP if the
client can successfully connect to our server. Employing the same sliding-window
technique defined in~\sectionref{sec:dns_filtering} in combination with the 90th
percentile threshold described above, we detect ESNI filtering in China, Russia,
and Iran.

Following a TLS client hello whose SNI field is encrypted, we observe RST
packets being injected by China's GFW to tear down connections destined for our
ESNI-supported web server. This observation aligns with previous anecdotal
reports that China has started filtering all ESNI traffic since July
2020~\cite{esnichina}.

Unlike the centralized ESNI filtering policy of China, Internet service
providers in Russia are known to implement blocking mechanisms independently in
a decentralized manner~\cite{Ramesh2020DecentralizedCA}. Among the networks
where we have VPs in Russia, we detect ESNI filtering in AS28890, AS52207, and
AS41754, where RST packets are injected to disrupt ESNI connections to our
website.

\sysname\ has also detected ESNI filtering from VPs in Iran's AS56402, AS31549,
and AS16322. However, since we did not have measurements from more than one
subnet in each of these ASes, we cannot conclude these cases as AS-level
filtering with high confidence.

\subsection{Network Filtering Circumvention}
\label{sec:circumvention}

\begin{table}[t]
  \centering
  \caption{Number of domains that could evade filtering as a result of domain
  name encryption employment. The filtering technique column indicates the
  number of domains that fail to evade filtering due to other filtering
  techniques (i.e., TCP packet injection), HTTP-only site, SNI-based filtering
  of domains without ESNI support, and server-side blocking.}
  \label{tab:crawl_results}
  \begin{tabular}{l@{\hskip 0.2in}c@{\hskip 0.2in}cc@{\hskip 0.15in}cc}
    \toprule
  \textbf{Country} & \multicolumn{1}{l}{\textbf{Circumvented/Total crawled}}{\hskip 0.2in} & \multicolumn{4}{l}{\textbf{Filtering technique}} \\
   & \multicolumn{1}{l}{} & TCP & HTTP & TLS & SS \\ \midrule
  China & 130/230 & 11 & 2 & 84 & 3 \\
  Russia & 53/56 & 1 & 1 & 1 & 0 \\
  Iran & 0/49 & 1 & 1 & 47 & 0 \\
  Indonesia & 93/98 & 2 & 2 & 0 & 1 \\
  India & 20/20 & 0 & 0 & 0 & 0 \\ \bottomrule
  \end{tabular}
  \end{table}

For the top five countries where most on-path DNS tampering was detected,
Table~\ref{tab:crawl_results} summarizes the number of domains that we (1) could
fetch by employing domain name encryption, thus evading network filtering, and
(2) could not fetch due to other filtering mechanisms at multiple layers of the
network stack.

Note that we focus our analysis on those domains tampered with by on-path
filtering systems (i.e., Table~\ref{tab:dns_table_public}), rather than those
blocked by local resolvers. This is because, instead of using domain name
encryption, simply changing to a public resolver is already sufficient to evade
filtering employed by a local resolver for some domains. This is the reason why
\sysname\ observes fewer filtered domains when querying public resolvers
(Table~\ref{tab:dns_table_public}) compared to local resolvers
(Table~\ref{tab:dns_table_local}).

Except for Iran, we could successfully unblock more than 50\% and 95\% of
filtered domains in China and other countries, respectively, where on-path DNS
filtering is heavily employed.

There are three reasons why some domains fail to evade filtering. First,
filtering systems can have several mechanisms deployed at different network
layers, as discussed in~\sectionref{sec:InternetFiltering}. Thus, domain name
encryption alone is not enough to cope with other filtering mechanisms. Second,
a few domains are still serving HTTP sites only, allowing straightforward
network interference. The third reason, which is also the main one, is because
many domains do not support ESNI (see Appendix~\ref{appendix:esni_prevalece}).
Domain name information of websites without ESNI support is still visible via
the TLS handshake and thus susceptible to SNI-based filtering.

Note that DNS-based and SNI-based filtering modules of China's GFW have been
shown to maintain different blocklists. Some domains, therefore, are filtered
via DNS tampering but not SNI-based interference~\cite{hoang:2019:measuringI2P}.
There are three domains that we could evade filtering in China but experience
server-side blocking.
\section{Related Work}
The adoption trend of domain name encryption technologies by major Internet
companies in the last couple of years has prompted several measurement studies
to examine how these new technologies are treated by Internet filtering
systems.

Basso~\cite{bassomeasuring} created a testing module to detect the blocking of
DoTH services for the OONI probe~\cite{filasto2012ooni}. The author analyzed
one-month data of measurements for 123 DoTH resolvers conducted by OONI
volunteers at three separate ASes in Kazakhstan, Iran, and China, finding that
the most frequently blocked DoTH resolvers belong to Cloudflare and Google.
While this study presents some preliminary insight into DoTH filtering,
\sysname\ conducts comprehensive measurements from more VPs at many network
locations over an extended period of time, providing a more complete view of how
filtering systems are treating domain name encryption technologies around the
globe. Specifically, the extensive and continuous measurements conducted by
\sysname\ have enabled us to discover exactly when a major filtering system like
the Great Firewall started blocking a domain name encryption protocol
(\sectionref{sec:dne_connectivity}).

Jin et al.~\cite{jin2021understanding} conducted a one-off measurement to
examine whether DoTH resolvers perform any DNS tampering themselves. Our system,
\sysname, is designed to also detect on-path filtering systems that interfere
with connections destined to DoTH servers. The authors also examined whether
encrypted DNS can help with bypassing Internet filtering by using commercial
VPNs running in data centers for testing. However, this has several drawbacks,
including VPN locations being falsified~\cite{Weinberg2018HowTC} and limited
visibility into residential networks that often have different filtering
policies~\cite{hoang:2019:measuringI2P}. The paper concludes that the
effectiveness of encrypted DNS in evading network interference varies by
country.

Chai et al.~\cite{chaiesni} study the adoption of ESNI and whether it can help
bypass Internet filtering in China. They found that 10.9\% of the Alexa top 1M
domains supported ESNI in 2018. By enabling ESNI in their web crawler, the
authors could unblock 66 websites filtered by the GFW based on
SNI~\cite{chaiesni}. Unfortunately, that has not gone unnoticed to China's
filtering systems. From July 2020, the GFW has been reported to block all ESNI
traffic~\cite{esnichina}. Our work complements this earlier work by verifying
the support of ESNI for all domains from TLD zone files, finding an increase to
almost 15\% of top 1M popular domains supporting ESNI
(\sectionref{sec:dne_connectivity}).
\section{Discussion}

From a technical perspective, it is obvious that domain name encryption
technologies can help to improve security and privacy for Internet users. Our
measurements, however, show mixed results when it comes to the resilience
of these technologies to Internet filtering. Specifically, while we found that
encrypting DNS resolutions could help evade DNS-based filtering for many
domains, almost half of the domains filtered in China and all domains filtered
in Iran could not evade filtering despite the use of DoH
(\sectionref{sec:circumvention}). This is primarily because the vast majority of
domains on the Internet do not have ESNI supported.

As a result, unless ESNI is universally deployed, DNS encryption alone is not
enough to resist Internet filtering. Moreover, filtering systems in China and
Russia have been blocking ESNI traffic because the collateral damage of this
blocking is not substantial enough, since only a small fraction of domains on the
Internet have ESNI supported. Even when more websites support ESNI, they should
be co-hosted instead of being hosted on separate IPs to increase the potential
collateral damage (if being blocked)~\cite{Hoang2020:CCR} and to avoid IP-based
blocking~\cite{Hoang2020:ASIACCS, Hoang2021:PoPETS}.

Another issue with DoTH is the chicken-and-egg problem of resolving the domains
of DoTH resolvers. Specifically, the domain of a DoTH resolver would still need
to be first resolved via an unencrypted DNS resolution. Although we did not
observe any major DNS-based filtering against the domains of DoTH resolvers, the
blocking cases of \texttt{ordns.he.net} and two Cloudflare DoH resolvers
in~\sectionref{sec:dns_filtering} show that this is a critical problem in the
current implementation of most DoTH resolvers. Although a client can instead use
a fixed IP of a DoTH resolver (e.g., \texttt{8.8.8.8} or \texttt{1.1.1.1}), this
setting is then susceptible to IP-based blocking unless the DoTH resolver's
identity is obfuscated similarly to our own DoH
resolver in~\sectionref{sec:measurements}. There have been studies demonstrating
the possibility of using machine learning models to detect and filter encrypted
DNS resolutions based on network traffic signatures~\cite{Csikor2021PrivacyOD,
Alenezi2021ClassifyingDT}, we however did not experience such blocking efforts.
This is evident by the fact that we could still use our private DoH resolver in
all countries where we have measurement vantage points.
\section{Limitations}

Prior work has shown that an advanced Internet filtering system such as the GFW
could block hundreds of thousands of domains~\cite{GFWatch}. Although it is
desirable to test as many domains as possible to obtain a more general view
about the filtering mechanisms used against various types of domains, we could
not test a large number of domains due to the short-lived nature of VPN Gate
vantage points (\sectionref{sec:testlist}). Another limitation of using VPN Gate
is that VPN endpoints often rewrite the packet header, taking away the
capability of using incremental IP time-to-live values to pinpoint the location of
filtering devices.
\section{Conclusion}

We present \sysname, a measurement system built on top of a distributed network
of vantage points, to examine the accessibility of domain name encryption
technologies and whether they are interfered with by filtering systems across
different network locations. Over a six-month period, \sysname\ conducted
315K measurements from more than 20K vantage points in 85 countries, detecting
blocking efforts against domain name encryption technologies in several
countries, including China, Russia, and Saudi Arabia.

Measuring the prevalence of ESNI adoption, we find that only 1.5--2.25\% of the
domains from TLD zone files have a valid ESNI key, indicating that ESNI has not
been widely adopted yet. Finally, to evaluate the efficacy of domain name
encryption in evading Internet filtering, we instrument a customized browser
with DoH and ESNI enabled to crawl a list of filtered domains detected by
\sysname. Except for network locations where SNI-based filtering is also
employed, we could unblock more than 55\% and 95\% of the blocked domains in
China and other countries where DNS-based filtering is employed.

\section*{Acknowledgments}

We would like to thank our shepherd, Gareth Tyson, and the anonymous reviewers
for their thorough feedback on earlier drafts of this paper. This research was
supported in part by the Open Technology Fund under an Information Controls
Fellowship. The opinions in this paper are those of the authors and do not
necessarily reflect the opinions of the sponsor.

\bibliographystyle{splncs04}
{\footnotesize
\bibliography{main}
}
\appendix
\vspace{-.3cm}
\section{DoTH Resolvers}
\label{appendix:doth_resolvers}

Table~\ref{tab:doth_mapping} indexes 71 DoTH resolvers publicly available at
the time of our study.

\begin{table*}[h!]
  \caption{
    The list of DoTH resolvers that is used in our measurement.}
  \label{tab:doth_mapping}
  \centering
  \resizebox{\columnwidth}{!}{
  \begin{tabular}{llllll}
  \toprule
  \textbf{Index} & \textbf{DoTH Servers}& \textbf{Index} & \textbf{DoTH Servers}& \textbf{Index} & \textbf{DoTH Servers} \\
  \midrule
  1  & 1dot1dot1dot1.cloudflare-dns.com& 25 & dns.switch.ch 			 	&  49 &  doh.xfinity.com          \\
  2  & cloudflare-dns.com 			& 26 & dns.twnic.tw 			 	&  50 &  family.cloudflare-dns.com           \\
  3  & dns10.quad9.net 				& 27 & dns-unfiltered.adguard.com	&  51 &  fi.doh.dns.snopyta.org           \\
  4  & dns11.quad9.net  				& 28 & doh-2.seby.io 				&  52 &  free.bravedns.com          \\
  5  & dns9.quad9.net 				& 29 & doh.applied-privacy.net 		&  53 &  jp.tiarap.org          \\
  6  & dns.aa.net.uk					& 30 & doh.centraleu.pi-dns.com		&  54 &  jp.tiar.app           \\
  7  & dns.adguard.com				& 31 & doh.cleanbrowsing.org 		&  55 &  mozilla.cloudflare-dns.com        \\
  8  & dns.alidns.com					& 32 & doh-de.blahdns.com 			&  56 &  odvr.nic.cz          \\
  9  & dns.containerpi.com 			& 33 & doh.dnslify.com				&  57 &  ordns.he.net          \\
  10 & dns.digitale-gesellschaft.ch 	& 34 & doh.dns.sb					&  58 &  resolver-eu.lelux.fi           \\
  11 & dns.dnshome.de  				& 35 & doh.eastas.pi-dns.com		&  59 &  security.cloudflare-dns.com           \\
  12 & dns.dns-over-https.com			& 36 & doh.eastau.pi-dns.com		&  60 &  1dot1dot1dot1.cloudflare-dns.com (DoT)           \\
  13 & dns.dnsoverhttps.net			& 37 & doh.eastus.pi-dns.com		&  61 &  adult-filter-dns.cleanbrowsing.org (DoT)           \\
  14 & dnses.alekberg.net				& 38 & doh.familyshield.opendns.com &  62 &  dns.adguard.com (DoT)           \\
  15 & dns-family.adguard.com			& 39 & doh.ffmuc.net				&  63 &  dns-family.adguard.com (DoT)           \\
  16 & dns.flatuslifir.is				& 40 & doh-fi.blahdns.com			&  64 &  dns.google (DoT)           \\
  17 & dnsforge.de					& 41 & doh-jp.blahdns.com			&  65 &  dns-nosec.quad9.net (DoT)           \\
  18 & dns.google						& 42 & doh.libredns.gr				&  66 &  dns.quad9.net (DoT)           \\
  19 & dns.hostux.net					& 43 & doh.northeu.pi-dns.com		&  67 &  dns-unfiltered.adguard.com (DoT)           \\
  20 & dnsnl.alekberg.net				& 44 & doh.opendns.com				&  68 &  dot.xfinity.com (DoT)            \\
  21 & dns-nosec.quad9.net			& 45 & doh.pi-dns.com				&  69 &  family-filter-dns.cleanbrowsing.org (DoT)           \\
  22 & dns-nyc.aaflalo.me				& 46 & doh.tiarap.org				&  70 &  one.one.one.one (DoT)           \\
  23 & dns.quad9.net					& 47 & doh.tiar.app					&  71 &  security-filter-dns.cleanbrowsing.org (DoT)          \\
  24 & dns.rubyfish.cn				& 48 & doh.westus.pi-dns.com		&   &             \\
  \bottomrule
  \end{tabular}}
  \end{table*}

\vspace{-.3cm}
\section{DNS Tampering Detection}
\label{appendix:dns_tampering}

To identify cases of DNS-based network interference, we employ the following
well-established consistency heuristics in the literature~\cite{filasto2012ooni,
pearce2017global, Satellite, niaki2020iclab}.

\myparab{Multiple responses with different ASes.} We receive multiple responses
for a DNS query that belong to different ASes. Previous studies have identified
cases where on-path filtering systems inject packets carrying false IP addresses
that often are publicly routable~\cite{GFWatch,niaki2020triplet,Anonymous2014a}.

\myparab{NXDomain or non-routable address.} We receive an NXDomain or
non-routable IP in response to a DNS query from a vantage point while receiving
a routable address from the majority of vantage points and our control node.

\myparab{Different responses from control and aggregate.} When a vantage point
receives a globally routable IP but different from the IP observed at the
control node. We first check whether they belong to the same AS. If both IPs are
under the same AS, this is due to the use of CDN and/or DNS-based load balancing
but not censorship. If the IP observed by the vantage point belongs to an AS
which is different from the response AS we observe at the control node and the
majority of other vantage points, this behavior indicates DNS interference by a
filtering system that aims to redirect the client to a different server (e.g.,
for displaying blockpages). However, there are also cases in which different
ASes are managed by large CDN providers (e.g., Akamai). We look up organization
information of those ASes to exclude cases where different response ASes belong
to the same organization to avoid false positives.

\vspace{-.3cm}
\section{AS-level DoTH Filtering}
\label{appendix:AS_doth_filtering}

Table~\ref{tab:AS_doth_filtering} shows the top five countries where most
connections to DoTH resolvers were interfered with. The DoTH server names are
indexed in Table~\ref{tab:doth_mapping}.

\begin{table}
  \caption{Top five countries where most AS-level DoTH filtering was detected. *
  indicate cases where both TCP and TLS handshakes were completed but we could
  not obtain the correct IP of our control domain being resolved.}
  \label{tab:AS_doth_filtering}
  \centering
  \resizebox*{!}{\textheight}{%
  \begin{tabular}{!{\vrule\vrule\vrule}c!{\vrule}c!{\vrule\vrule\vrule}c!{\vrule}c!{\vrule}c!{\vrule}c!{\vrule}c!{\vrule}c!{\vrule\vrule\vrule}c!{\vrule}c!{\vrule}c!{\vrule}c!{\vrule\vrule\vrule}c!{\vrule}c!{\vrule}c!{\vrule\vrule\vrule}c!{\vrule}c!{\vrule}c!{\vrule\vrule\vrule}c!{\vrule}c!{\vrule}c!{\vrule\vrule\vrule}c!{\vrule\vrule\vrule}}
  \hline
  \multicolumn{2}{!{\vrule\vrule\vrule}c!{\vrule\vrule\vrule}}{Country}           & \multicolumn{6}{c!{\vrule\vrule\vrule}}{China}                                                                                                                                                                                                                                                      & \multicolumn{4}{c!{\vrule\vrule\vrule}}{United States}                                                                                                                                              & \multicolumn{3}{c!{\vrule\vrule\vrule}}{S. Korea}                                                                                               & \multicolumn{3}{c!{\vrule\vrule\vrule}}{Singapore}                                                                                                 & \multicolumn{3}{c!{\vrule\vrule\vrule}}{Saudi Arabia}                                                                                              & Iran                                             \\
  \hline
  \multicolumn{2}{!{\vrule\vrule\vrule}c!{\vrule\vrule\vrule}}{ASN}               & \rotatebox[origin=c]{90}{4134}              & \rotatebox[origin=c]{90}{4837}                 & \rotatebox[origin=c]{90}{9808}                 & \rotatebox[origin=c]{90}{37963}                & \rotatebox[origin=c]{90}{45090}                & \rotatebox[origin=c]{90}{140314}                & \rotatebox[origin=c]{90}{7155}                & \rotatebox[origin=c]{90}{20473}                & \rotatebox[origin=c]{90}{31898}                & \rotatebox[origin=c]{90}{36352} & \rotatebox[origin=c]{90}{17870} & \rotatebox[origin=c]{90}{20473} & \rotatebox[origin=c]{90}{38121} & \rotatebox[origin=c]{90}{14061} & \rotatebox[origin=c]{90}{20473} & \rotatebox[origin=c]{90}{55430} & \rotatebox[origin=c]{90}{25019} & \rotatebox[origin=c]{90}{35819} & \rotatebox[origin=c]{90}{35753} & \rotatebox[origin=c]{90}{58224}  \\
  \hline\hline
  \multirow{37}{*}{\rotatebox[origin=c]{90}{Index of blocked DoH resolvers}}& 1   & ~    & ~    & X    & ~     & ~     & ~          & ~    & ~     & ~     & ~                                & ~     & ~     & X                                     & ~     & ~     & ~                                   & X     & X     & X                                      & X                                               \\
  \cline{2-22}
                                  & 2                                             & ~    & ~    & X    & ~     & ~     & X          & ~    & ~     & ~     & ~                                & ~     & ~     & ~                                     & ~     & ~     & ~                                   & X     & X     & X                                      & ~                                               \\ 
  \cline{2-22}
                                  & 4                                             & ~    & ~    & ~    & ~     & ~     & ~          & ~    & ~     & ~     & ~                                & ~     & ~     & ~                                     & ~     & ~     & ~                                   & ~     & ~     & ~                                      & X                                               \\ 
  \cline{2-22}
                                  & 5                                             & ~    & ~    & ~    & ~     & ~     & X          & ~    & ~     & ~     & ~                                & ~     & ~     & ~                                     & ~     & ~     & ~                                   & ~     & ~     & ~                                      & ~                                               \\ 
  \cline{2-22}
                                  & 6                                             & ~    & ~    & X    & ~     & ~     & X          & ~    & ~     & ~     & ~                                & ~     & ~     & ~                                     & ~     & ~     & ~                                   & ~     & ~     & ~                                      & ~                                               \\ 
  \cline{2-22}
                                  & 9                                             & ~    & ~    & X    & ~     & ~     & X          & ~    & ~     & ~     & ~                                & ~     & ~     & ~                                     & ~     & ~     & ~                                   & ~     & ~     & ~                                      & ~                                               \\ 
  \cline{2-22}
                                  & 10                                            & ~    & ~    & X    & ~     & ~     & X          & ~    & ~     & ~     & ~                                & ~     & ~     & ~                                     & ~     & ~     & ~                                   & ~     & ~     & ~                                      & ~                                               \\
  \cline{2-22}
                                  & 11                                            & ~    & ~    & ~    & ~     & ~     & X          & ~    & ~     & ~     & ~                                & ~     & ~     & ~                                     & ~     & ~     & ~                                   & ~     & ~     & ~                                      & ~                                               \\ 
  \cline{2-22}
                                  & 13                                            & ~    & ~    & X    & ~     & ~     & ~          & ~    & ~     & ~     & ~                                & ~     & ~     & ~                                     & ~     & ~     & ~                                   & ~     & ~     & ~                                      & ~                                               \\ 
  \cline{2-22}
                                  & 14                                            & ~    & ~    & X    & ~     & ~     & X          & ~    & ~     & ~     & ~                                & ~     & ~     & ~                                     & ~     & ~     & ~                                   & ~     & ~     & ~                                      & ~                                               \\ 
  \cline{2-22}
                                  & 16                                            & ~    & ~    & X    & ~     & ~     & X          & ~    & ~     & ~     & ~                                & ~     & ~     & ~                                     & ~     & ~     & ~                                   & ~     & ~     & ~                                      & ~                                               \\ 
  \cline{2-22}
                                  & 18                                            & X    & X    & X    & X     & X     & X          & ~    & ~     & ~     & ~                                & ~     & ~     & ~                                     & ~     & ~     & ~                                   & ~     & ~     & ~                                      & X                                               \\ 
  \cline{2-22}
                                  & 19                                            & ~    & ~    & X    & ~     & ~     & X          & ~    & ~     & ~     & ~                                & ~     & ~     & ~                                     & ~     & ~     & ~                                   & ~     & ~     & ~                                      & ~                                               \\ 
  \cline{2-22}
                                  & 20                                            & ~    & ~    & X    & ~     & ~     & X          & ~    & ~     & ~     & ~                                & ~     & ~     & ~                                     & ~     & ~     & ~                                   & ~     & ~     & ~                                      & ~                                               \\ 
  \cline{2-22}
                                  & 22                                            & ~    & ~    & X    & ~     & ~     & ~          & ~    & ~     & ~     & ~                                & ~     & ~     & ~                                     & ~     & ~     & ~                                   & ~     & ~     & ~                                      & ~                                               \\ 
  \cline{2-22}
                                  & 23                                            & ~    & ~    & X    & ~     & ~     & ~          & ~    & ~     & ~     & ~                                & ~     & ~     & ~                                     & ~     & ~     & ~                                   & ~     & ~     & ~                                      & X                                               \\ 
  \cline{2-22}
                                  & 25                                            & ~    & ~    & X    & ~     & ~     & ~          & ~    & ~     & ~     & ~                                & ~     & ~     & ~                                     & ~     & ~     & ~                                   & ~     & ~     & ~                                      & ~                                               \\ 
  \cline{2-22}
                                  & 26                                            & ~    & ~    & ~    & X     & ~     & X          & ~    & ~     & ~     & ~                                & ~     & ~     & ~                                     & ~     & ~     & ~                                   & ~     & ~     & ~                                      & ~                                               \\ 
  \cline{2-22}
                                  & 27                                            & ~    & ~    & ~    & ~     & ~     & X          & ~    & ~     & ~     & ~                                & ~     & ~     & ~                                     & ~     & ~     & ~                                   & ~     & ~     & ~                                      & ~                                               \\ 
  \cline{2-22}
                                  & 28                                            & ~    & ~    & X    & ~     & ~     & X          & ~    & ~     & ~     & ~                                & ~     & ~     & ~                                     & ~     & ~     & ~                                   & ~     & ~     & ~                                      & ~                                               \\ 
  \cline{2-22}
                                  & 32                                            & ~    & ~    & X    & ~     & ~     & ~          & ~    & ~     & ~     & ~                                & ~     & ~     & ~                                     & ~     & ~     & ~                                   & ~     & ~     & ~                                      & ~                                               \\ 
  \cline{2-22}
                                  & 33                                            & ~    & ~    & X    & X     & X     & ~          & ~    & ~     & ~     & ~                                & ~     & ~     & ~                                     & ~     & ~     & ~                                   & ~     & ~     & ~                                      & ~                                               \\ 
  \cline{2-22}
                                  & 34                                            & ~    & ~    & X    & ~     & ~     & X          & ~    & ~     & ~     & ~                                & ~     & ~     & ~                                     & ~     & ~     & ~                                   & ~     & ~     & ~                                      & ~                                               \\ 
  \cline{2-22}
                                  & 37                                            & ~    & ~    & X    & ~     & ~     & X          & ~    & ~     & ~     & ~                                & ~     & ~     & ~                                     & ~     & ~     & ~                                   & ~     & ~     & ~                                      & ~                                               \\ 
  \cline{2-22}
                                  & 38                                            & ~    & ~    & X    & X     & X     & X          & X    & ~     & ~     & X                                & X     & ~     & ~                                     & X     & X     & X                                   & ~     & ~     & ~                                      & X                                               \\ 
  \cline{2-22}
                                  & 39                                            & ~    & ~    & X    & ~     & ~     & X          & ~    & ~     & ~     & ~                                & ~     & ~     & ~                                     & ~     & ~     & ~                                   & ~     & ~     & ~                                      & ~                                               \\ 
  \cline{2-22}
                                  & 40                                            & ~    & ~    & X    & X     & X     & X          & ~    & ~     & ~     & ~                                & ~     & ~     & ~                                     & ~     & ~     & ~                                   & ~     & ~     & ~                                      & ~                                               \\ 
  \cline{2-22}
                                  & 41                                            & ~    & ~    & X    & X     & X     & ~          & ~    & ~     & ~     & ~                                & ~     & ~     & ~                                     & ~     & ~     & ~                                   & ~     & ~     & ~                                      & ~                                               \\ 
  \cline{2-22}
                                  & 42                                            & ~    & ~    & X    & ~     & ~     & X          & ~    & ~     & ~     & ~                                & ~     & ~     & ~                                     & ~     & ~     & ~                                   & ~     & ~     & ~                                      & ~                                               \\ 
  \cline{2-22}
                                  & 44                                            & ~    & ~    & X    & X     & X     & X          & X    & ~     & X     & X                                & X     & X     & ~                                     & X     & X     & X                                   & ~     & ~     & ~                                      & ~                                               \\ 
  \cline{2-22}
                                  & 45                                            & X    & X    & ~    & X     & X     & ~          & ~    & ~     & ~     & ~                                & ~     & ~     & ~                                     & ~     & ~     & ~                                   & ~     & ~     & ~                                      & ~                                               \\ 
  \cline{2-22}
                                  & 47                                            & ~    & ~    & X    & ~     & ~     & X          & ~    & X     & ~     & ~                                & ~     & ~     & ~                                     & ~     & X     & ~                                   & ~     & ~     & ~                                      & ~                                               \\ 
  \cline{2-22}
                                  & 48                                            & ~    & ~    & X    & ~     & ~     & X          & ~    & ~     & ~     & ~                                & ~     & ~     & ~                                     & ~     & ~     & ~                                   & ~     & ~     & ~                                      & ~                                               \\ 
  \cline{2-22}
                                  & 50                                            & ~    & ~    & ~    & ~     & ~     & ~          & ~    & ~     & ~     & ~                                & ~     & ~     & X                                     & ~     & ~     & ~                                   & X     & X     & X                                      & X                                               \\ 
  \cline{2-22}
                                  & 51                                            & ~    & ~    & X    & ~     & ~     & X          & ~    & ~     & ~     & ~                                & ~     & ~     & ~                                     & ~     & ~     & ~                                   & ~     & ~     & ~                                      & ~                                               \\ 
  \cline{2-22}
                                  & 54                                            & ~    & ~    & ~    & ~     & ~     & X          & ~    & X     & ~     & ~                                & ~     & ~     & ~                                     & ~     & X     & ~                                   & ~     & ~     & ~                                      & ~                                               \\ 
  \cline{2-22}
                                  & 55                                            & ~    & ~    & X    & X     & X     & X          & ~    & ~     & ~     & ~                                & ~     & ~     & ~                                     & ~     & ~     & ~                                   & X     & X     & X                                      & ~                                               \\ 
  \cline{2-22}
                                  & 59                                            & ~    & ~    & ~    & ~     & ~     & ~          & ~    & ~     & ~     & ~                                & ~     & ~     & X                                     & ~     & ~     & ~                                   & X     & X     & X                                      & X                                               \\ 
  \hline\hline
  \multirow{12}{*}{\rotatebox[origin=c]{90}{Index of blocked DoT resolvers}} & 60 & ~    & ~    & X    & ~     & ~     & X          & ~    & ~     & ~     & ~                                & ~     & ~     & X                                     & ~     & ~     & ~                                   & ~     & ~     & ~                                      & ~                                               \\
  \cline{2-22}
                                  & 61                                            & ~    & ~    & X    & ~     & ~     & X          & ~    & ~     & ~     & ~                                & ~     & ~     & ~                                     & ~     & ~     & ~                                   & ~     & ~     & ~                                      & ~                                               \\ 
  \cline{2-22}
                                  & 62                                            & ~    & ~    & X    & ~     & ~     & X          & ~    & ~     & ~     & ~                                & ~     & ~     & ~                                     & ~     & ~     & ~                                   & ~     & ~     & ~                                      & ~                                               \\ 
  \cline{2-22}
                                  & 63                                            & ~    & ~    & X    & ~     & ~     & X          & ~    & ~     & ~     & ~                                & ~     & ~     & ~                                     & ~     & ~     & ~                                   & ~     & ~     & ~                                      & ~                                               \\ 
  \cline{2-22}
                                  & 64                                            & ~    & ~    & ~    & ~     & ~     & ~          & ~    & ~     & ~     & ~                                & ~     & ~     & ~                                     & ~     & ~     & ~                                   & ~     & ~     & ~                                      & X                                               \\ 
  \cline{2-22}
                                  & 65                                            & ~    & ~    & X    & ~     & ~     & X          & ~    & ~     & ~     & ~                                & ~     & ~     & ~                                     & ~     & ~     & ~                                   & ~     & ~     & ~                                      & ~                                               \\ 
  \cline{2-22}
                                  & 66                                            & ~    & ~    & X    & ~     & ~     & X          & ~    & ~     & ~     & ~                                & ~     & ~     & ~                                     & ~     & ~     & ~                                   & ~     & ~     & ~                                      & ~                                               \\ 
  \cline{2-22}
                                  & 67                                            & ~    & ~    & X    & ~     & ~     & X          & ~    & ~     & ~     & ~                                & ~     & ~     & ~                                     & ~     & ~     & ~                                   & ~     & ~     & ~                                      & ~                                               \\ 
  \cline{2-22}
                                  & 68                                            & ~    & ~    & X    & ~     & ~     & X          & ~    & ~     & ~     & ~                                & ~     & ~     & ~                                     & ~     & ~     & ~                                   & ~     & ~     & ~                                      & ~                                               \\ 
  \cline{2-22}
                                  & 69                                            & ~    & ~    & X    & ~     & ~     & X          & ~    & ~     & ~     & ~                                & ~     & ~     & ~                                     & ~     & ~     & ~                                   & ~     & ~     & ~                                      & ~                                               \\ 
  \cline{2-22}
                                  & 70                                            & ~    & ~    & X    & ~     & ~     & X          & ~    & ~     & ~     & ~                                & ~     & ~     & X                                     & ~     & ~     & ~                                   & ~     & ~     & ~                                      & ~                                               \\ 
  \cline{2-22}
                                  & 71                                            & ~    & ~    & X    & ~     & ~     & X          & ~    & ~     & ~     & ~                                & ~     & ~     & ~                                     & ~     & ~     & ~                                   & ~     & ~     & ~                                      & ~                                               \\ 
  \hline\hline
  Block                           & TCP                                           & 98.3                                          & 97.8                                             & 44.8                                            & 4.8                                             & 95.3                                             & 96.2                                              & 0*                                            & 100                                             & 0*                                             & 0*                                             & 100                                             & 0*                                             & 100                                             & 0*                                             & 50                                             & 0*                                             & 0                                             & 0                                             & 0                                             & 10                                              \\
  \cline{2-22}
  (\%)                            & TLS                                           & 1.7                                          &2.2                                             & 55.2                                              & 95.2                                               & 4.7                                               & 3.8                                                & 0*                                              & 0                                               & 0*                                              & 0*                                               & 0                                               & 0*                                               & 0                                              & 0*                                               & 50                                               & 0*                                              & 100                                               & 100                                               & 100                                               & 90                                                \\
  \hline
  \end{tabular}
  }
  \arrayrulecolor{black}
  \end{table}

\section{ESNI Prevalence}
\label{appendix:esni_prevalece}

Over the course of our measurement period, we frequently query for ESNI TXT
records of more than 350M domains from TLD zone files~\cite{ICANN}. Only
3\%--4.5\% of domains respond to our ESNI TXT queries. And, only 48--51\% of
these TXT records have a valid ESNI key format defined in the Internet
drafts~\cite{tlsdraft02, tlsdraft05}. Analyzing the key lengths of all ESNI TXT
records obtained, we find that the majority of them have 92 characters. These
ESNI-supported domains are hosted by Cloudflare, which is the only Internet
company supporting ESNI to the best of our knowledge. For domains whose ESNI TXT
records that do not have a correct ESNI key format, we find that their
authoritative nameservers are configured with a wildcard setup (i.e.,
\texttt{\mbox{*}.example.com}), thus responding to our ESNI TXT query for
\texttt{\_esni.example.com} despite not having an actual ESNI key. To that end,
only around 1.5\%--2.25\% of domains on the Internet have ESNI supported.

\end{document}